\documentclass[10pt,conference]{IEEEtran}
\IEEEoverridecommandlockouts

\usepackage[utf8]{inputenc}
\usepackage[T1]{fontenc}
\usepackage{xcolor}
\usepackage{microtype}
\usepackage{xspace}
\usepackage{graphicx}
\usepackage{subfigure}
\usepackage{booktabs}
\usepackage{balance}
\usepackage{amsmath}
\usepackage[linewidth=1pt]{mdframed}
\usepackage{enumitem}
\usepackage{multirow}
\usepackage{makecell}
\usepackage{amssymb}
\usepackage{romannum}
\usepackage{threeparttable}
\usepackage[most]{tcolorbox}
\usepackage{marvosym}

\newcommand{\etal}{{\em et al.}\xspace}
\newcommand{\ie}{{\em i.e.},\xspace}
\newcommand{\eg}{{\em e.g.},\xspace}

\newcommand{\wechat}{\textsl{WeChat}\xspace}
\newcommand{\work}{\textsl{WeChat-Work}\xspace}
\newcommand{\info}{\textsl{WeChat-Info}\xspace}
\newcommand{\pay}{\textsl{WeChat-Pay}\xspace}
\newcommand{\game}{\textsl{WeChat-Game}\xspace}
\newcommand{\weread}{\textsl{WeChat-Reading}\xspace}

\begin{document}

\title{Can User Feedback Help Issue Detection? \\ An Empirical Study on a One-billion-user \\ Online Service System}

\author{
\IEEEauthorblockN{Shuyao Jiang\textsuperscript{*}, Jiazhen Gu\textsuperscript{*}, Wujie Zheng\textsuperscript{\dag,\ddag}, Yangfan Zhou\textsuperscript{\dag,\Letter}\thanks{\textsuperscript{\Letter} Yangfan Zhou is the corresponding author.} and Michael R. Lyu\textsuperscript{*}}
\IEEEauthorblockA{\textsuperscript{*} \textit{Department of Computer Science and Engineering, The Chinese University of Hong Kong, Hong Kong, China}}
\IEEEauthorblockA{\textsuperscript{\dag} \textit{College of Computer Science and Artificial Intelligence, Fudan University, Shanghai, China}}
\IEEEauthorblockA{\textsuperscript{\ddag} \textit{Tencent Inc., Shenzhen, China}}
\IEEEauthorblockA{syjiang21@cse.cuhk.edu.hk, jiazhengu@cuhk.edu.hk, wujiezheng@tencent.com, zyf@fudan.edu.cn, lyu@cse.cuhk.edu.hk}
}

\maketitle

\begin{abstract}
\textit{Background:} It has long been suggested that user feedback, typically written in natural language by end-users, can help issue detection. However, for large-scale online service systems that receive a tremendous amount of feedback, it remains a challenging task to identify severe issues from user feedback. 

\textit{Aims:} To develop a better feedback-based issue detection approach, it is crucial first to gain a comprehensive understanding of the characteristics of user feedback in real production systems.

\textit{Method:} In this paper, we conduct an empirical study on 50,378,766 user feedback items from six real-world services in a one-billion-user online service system. We first study what users provide in their feedback. We then examine whether certain features of feedback items can be good indicators of severe issues. Finally, we investigate whether adopting machine learning techniques to analyze user feedback is reasonable. 

\textit{Results:} Our results show that a large proportion of user feedback provides irrelevant information about system issues. As a result, it is crucial to filter out issue-irrelevant information when processing user feedback. 
Moreover, we find severe issues that cannot be easily detected based solely on user feedback characteristics. 
Finally, we find that the distributions of the feedback topics in different time intervals are similar. This confirms that designing machine learning-based approaches is a viable direction for better analyzing user feedback. 

\textit{Conclusions:} We consider that our findings can serve as an empirical foundation for feedback-based issue detection in large-scale service systems, which sheds light on the design and implementation of practical issue detection approaches.
\end{abstract}


\renewcommand{\thefootnote}{}
\footnotetext{979-8-3315-9147-2/25/\$31.00 ©2025 IEEE}

\section{Introduction}

User feedback on software systems, typically written in natural language by end-users, reflects the instant user experience with the systems. It has long been suggested that user feedback is an important data resource that can assist software development and maintenance \cite{DBLP:conf/re/PaganoM13, DBLP:conf/icse/ChenLHXZ14, DBLP:conf/wsdm/NguyenLT15, DBLP:conf/icse/VillarroelBROP16}. 

Issues, \eg software faults, are inevitable in software systems~\cite{lyu1996handbook}. Hence, large-scale cloud services, especially those of major providers (\eg Facebook and Microsoft) typically provide a convenient interface to collect user feedback on the bad experiences users have encountered \cite{DBLP:conf/kbse/ZhengLZLZD19}. 
Since such services serve millions of users around the world on a 24/7 basis, they can receive tremendous user feedback every day. 

Recently, extensive research work has proposed conducting issue analysis based on such user feedback.
Example proposals include those 
for offline, postmortem analysis, \eg mining user requirements \cite{DBLP:conf/icse/PaganoB13,DBLP:conf/issre/GaoWHZZL15,DBLP:conf/www/LuizVAMSCGR18} and directing software upgrading~\cite{DBLP:journals/software/KhalidSNH15,DBLP:conf/icse/GaoZLK18,DBLP:journals/jss/PalombaVBOPPL18}. 
Online issue detection based on user feedback is also shown as a promising solution \cite{DBLP:conf/kbse/ZhengLZLZD19,DBLP:conf/icse/GaoZD0ZLK19}.

However, in current industry practice, user feedback may contain various issues, including those related to severe service defects and those irrelevant to the target service. 
It is infeasible to manually inspect the issues reported from user feedback.
Whereas, effective automatic detection of severe issues from user feedback still remains a challenging task \cite{DBLP:conf/kbse/ZhengLZLZD19}. 
As a result, in current industry practice, it is quite common that many feedback items in online service systems are just stored, without being further analyzed to accurately identify the issues they describe.
Towards a better feedback-based issue detection approach, it is critical 
to first obtain a comprehensive understanding of the characteristics of real-world user feedback. However, current literature still lacks such an understanding. This work aims to bridge this gap through a large-scale empirical study of a one-billion-user online service system. 

In particular, recent research assumes user feedback is mostly on reporting 
issues, and proposes to conduct issue detection accordingly \cite{DBLP:conf/kbse/ZhengLZLZD19, DBLP:conf/icse/GaoZD0ZLK19}. We verify this consideration by studying
what users will provide in their feedback in real-world production services.
Moreover, the reported issues in user feedback are not equally important in improving service quality. Many approaches have proposed that issues can be ranked according to their severity inferred based on the characteristics of their corresponding feedback items \cite{DBLP:conf/icse/GaoZLK18, DBLP:conf/icse/GaoZD0ZLK19}. Manual inspection efforts can thus be saved by directing the inspection focus to severe issues. We therefore examine whether some popular characteristics (\eg sentiment, text length, and history data) of user feedback items can be good indicators of issue severity. Finally, current approaches largely rely on machine learning techniques \cite{DBLP:books/daglib/0087929} to analyze user feedback \cite{DBLP:conf/icse/GaoZLK18}. 
The prerequisite is that the statistical characteristics of the feedback are stable over time, which can then be learned based on historical data.
We, therefore, investigate whether this consideration holds for user feedback. 
 
To this end, we conduct an empirical study of {\em all} the user feedback (in total, 50,378,766 items) collected in one year for six real-world services in a one-billion-user online service system. We find that there exists much irrelevant information in the feedback, where only a small proportion (which may be lower than $30\%$) of the feedback items are issue-relevant. A preprocessing mechanism to filter irrelevant information is therefore necessary. Moreover, it is reasonable that the severity of the reported issue is related to the number of its corresponding feedback items, as suggested in current practice~\cite{DBLP:conf/kbse/ZhengLZLZD19}. However, we also observe exceptions, where some severe issues are only reported by a few users. The text-based features (\ie sentiment, text length) of individual feedback items, in contrast, are not good criteria for identifying potential severe issues. But, proper user modeling can, to some extent, help determine whether the issue a user reports is severe. Finally, we find that the distributions of the feedback topics in different time intervals are largely similar, which shows that machine learning is a viable means to 
analyze user feedback. 

In summary, this work makes the following contributions:
\begin{itemize}
    \item Our study serves as the first attempt to provide a comprehensive understanding of real-world user feedback in large-scale online service systems.
    \item We empirically analyze 50 million feedback items from a billion-user system, revealing that traditional text features poorly predict issue severity, while user behavior and temporal patterns offer more reliable signals.
    \item Our findings provide an empirical foundation for feedback-based online issue detection methods, particularly in filtering noise, prioritizing severe issues, and validating the feasibility of machine learning for feedback analysis.
\end{itemize}

The data used in this paper are proprietary and cannot be publicly released due to confidentiality restrictions. The methodology is described in sufficient detail to ensure verifiability.

\section{Background and Motivation}
\label{sec:background}
 

Our target service system is one for hosting services for a popular social media platform, \textit{WeChat}~\cite{wechat}. 
It has penetrated many aspects of users' social lives, including work, entertainment, and business activities. It consists of tens of online services, serving over a billion users. Such services are quite diverse 
in functionality, including those for general social media applications (\eg instant messaging and update sharing), online payment, and social games. 
The platform uses a single sign-on (SSO) policy, allowing users to access all services conveniently. As a result, all the services have a tremendous number of concurrent users on a 24/7 basis. 
 
\begin{figure}[t!]
    \centering
    \includegraphics[width=\linewidth]{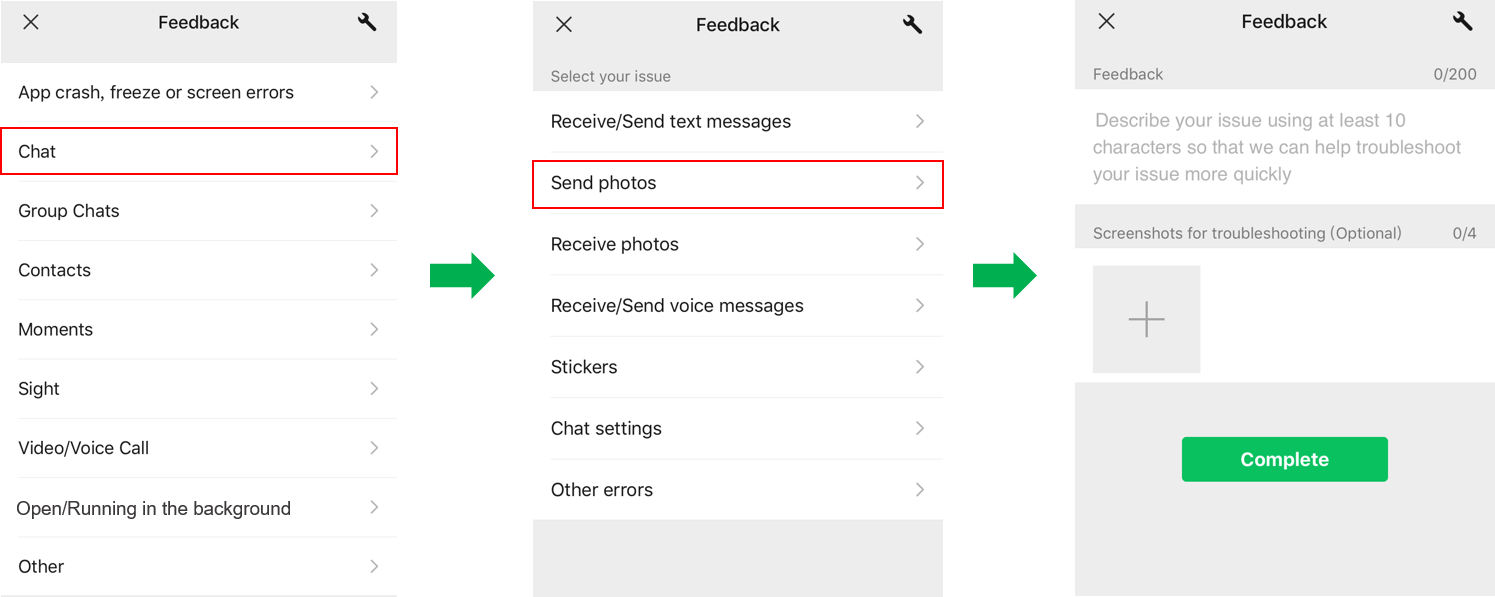}
     
    %
    \caption{A typical user feedback interface.}
    \label{fig:feedback_interface}
\end{figure}

Each service of the platform provides a feedback interface to its users.
Figure \ref{fig:feedback_interface} shows a typical tree-structured user feedback interface, which allows users to write text comments and optionally upload pictures.
The user may first need to choose the category of the issue to be reported. For example, the user may select the category ``Chat'' to indicate she intends to report an issue on the instance messenger. She may select ``Send photos'' in the next interface as she intends to report something related to sending a photo, and finally reach the interface of detailed feedback.

\begin{figure} 
\centering
    \begin{mdframed}
        \small \textbf{Item 1:} I cannot send and receive any message in the chat \\
        \small \qquad group, please solve the problem. \\
        \small \textbf{Item 2:} Why does the app crash when sending a video? \\
        \small \textbf{Item 3:} I cannot add new friends, what should I do? 
    \end{mdframed}
    \caption{Example user feedback items.}
    \label{fig:feedback_example}
\end{figure}

In a user feedback item, a user typically reports the issue she encounters when using a certain functionality of a service in natural language. Figure \ref{fig:feedback_example} shows some example feedback texts for the instant messenger service. 
Since such user feedback interfaces are intentionally designed to collect bad user experiences, the feedback items typically contain the  
complaints and the issues users have encountered when using the services. 
For example, the first feedback item in Figure \ref{fig:feedback_example} 
indicates that the user has encountered difficulties sending or receiving instant messages.

Due to the extremely large number of users, the platform can collect tremendous amounts of user feedback, typically around two million, on a daily basis. Naturally, one may suggest that such a huge amount of feedback items can facilitate fault detection and fault localization of the services. 
Actually, user feedback has long been suggested as a profitable source of fault detection, typically in a postmortem analysis manner \cite{DBLP:conf/issre/GaoWHZZL15}. Recent work also proposes to conduct online issue detection with such feedback \cite{DBLP:conf/kbse/ZhengLZLZD19}. 

To conduct issue detection based on user feedback, one prerequisite is that the feedback items should clearly indicate issues. However, we still lack a good understanding of 
what users will provide in their feedback in real-world scenarios. Therefore, we have the following research question: 
 
\textit{RQ1: What is the proportion of feedback items that report issues in real-world online services?}

By answering this research question, we can examine whether it is feasible to conduct issue detection based on user feedback. Then, if there is a large volume of issue-relevant feedback items in real-world online services, an instant consideration is how to conduct issue detection in user feedback.
Manual inspection of every feedback item may not be practical if tremendous issues are collected during service runtime. In industry practice, it is common that most user feedback issues are neglected to save human inspection efforts~\cite{DBLP:conf/kbse/ZhengLZLZD19}. A straightforward consideration is that issues can be ranked according to their severity. Top-ranked issues deserve further manual inspection. 
This is a basic notion to many issue-mining approaches \cite{DBLP:conf/kbse/ZhengLZLZD19,DBLP:conf/icse/GaoZLK18, DBLP:conf/icse/GaoZD0ZLK19}. Therefore, we are motivated to investigate the following research question: 

\textit{RQ2: Can the number of feedback items on an issue indicate how severe the issue is?}

Another consideration is whether a feedback item itself has provided enough information on the severity of its reported issue, as follows:

\textit{RQ3: Can certain features of a feedback item indicate the severity of its reported issue?  }

Answering RQ2 and RQ3 can provide potential directions for better mining system issues from user feedback.  Finally, as natural language processing with machine learning is widely adopted in analyzing user feedback items, we consider whether machine learning is suitable for analyzing the data (\ie user feedback). Since machine learning approaches generally learn a statistical model based on user feedback items, one requirement is that the statistical characteristics of the feedback are stable over time. Otherwise, the model may be misled by historical data and produce bad results when applied to future data. In other words, the prerequisite of adopting machine learning is that the training data (\ie historical feedback) shares a similar distribution to the future, unknown data (\ie future feedback). Therefore, the following research question is of interest: 

\textit{RQ4: Will the topics of user feedback change significantly over time?}

Issue detection is a critical task in online service maintenance. Answering these four fundamental questions can provide insights into the design of practical issue detection methods. 
Our work serves as a first attempt to answer these questions with an empirical study on a real-world, large-scale online service platform that serves over one billion users. 
\section{Empirical Study}\label{sec:findings}

In this section, we provide our empirical study on the user feedback items collected during a one-year period of six representative services. We aim to answer the above four research questions by analyzing the feedback items. Next, we first discuss how we collect data, followed by elaborating on our study on each research question. 
 
\subsection{Data Collection} 

Our target social media platform provides tens of online services to end users. As our aim is to study the characteristics of user feedback, we choose our target services among those that collect the most user feedback items. In addition, we also take diverse service domains and usage scenarios into consideration. In this way, we select six services from our target social media platform as our study objects. 
Table \ref{table:services} provides information about the six target services. 

\begin{table}[t!]
\scriptsize
\centering
\caption{Six services of our target social media platform.}
\resizebox{0.9\linewidth}{!}{
\begin{tabular}{ l  c  c }
 \toprule
 \textbf{Service} &  \textbf{Functionality} & \textbf{User Scale} \\
 \midrule
 \textsl{WeChat} & Instant Messaging & $10^{9}$\\
 \textsl{WeChat-Work} & Workplace Communication & $10^{8}$ \\
 \textsl{WeChat-info} & Content Subscription & $10^{8}$ \\
 \textsl{WeChat-Pay} & Mobile Payment & $10^{8}$ \\ 
 \textsl{WeChat-Game} & Mobile Game & $10^{8}$ \\ 
 \textsl{WeChat-Reading} & Reading & $10^{8}$ \\ 
 \bottomrule
\end{tabular}}
\label{table:services}
\end{table} 

Specifically, \wechat is a service that provides instant messaging functionalities. Users typically use it to chat with friends and share updates. It is the core service of the social media platform, which serves over one billion daily active users.  
\work provides a workplace communication service for enterprise users, through which users can communicate with colleagues and share or transfer documents. Currently, \work has been serving millions of organizations with over one hundred million users. 
\info is a content subscription service, where users can subscribe to the organization or individual official accounts. Users can also create their own official accounts in \info to publish their articles. \pay provides a mobile payment service for users to conveniently conduct online or face-to-face purchases and transfer money. \game is an entertainment service that provides a mobile game platform, hosting a variety of games. \weread is a reading service that allows users to buy, download, and read e-books on their mobile devices.

These six services are representative of large-scale online services. They are serving tremendous concurrent users with diverse functionalities, which cover different aspects of users' social lives, including those for daily and workplace social interactions (\ie~\wechat, \work, and \info), entertainment (\ie~\game and \weread), and task-oriented facility (\ie~\pay). 

Each of these services receives an average of more than 20,000 feedback items per day.
To better comprehend the characteristics of user feedback and avoid sampling bias, we do not take a sampling approach to collect feedback items. Instead, we collect the {\em entire set} of all user feedback items in one year. 
In total, we collect 50,378,766 feedback items.

\subsection{RQ1: What is the proportion of feedback items that report issues in real-world online services?}
\label{finding:rq1}

Large-scale online service systems can receive massive user feedback on a daily basis, due to their huge user numbers. 
For example, our field study finds that in our target platform, each target service typically receives several thousand to tens of thousands of feedback items per day.
Since the user feedback mechanism is designed to allow users to report their 
complaints and the issues encountered, an instant approach to exploit such feedback is to analyze the contents and accordingly mine the issues (\ie the corresponding system faults), as suggested by some recent work \cite{DBLP:conf/issre/GaoWHZZL15, DBLP:conf/icse/GaoZLK18}. 

But before designing sophisticated methods to analyze the feedback contents, the prerequisite to mine the issues from the user feedback items is that they are in nature those to report issues. Hence, we are motivated to examine whether this prerequisite holds in real-world scenarios. 

We first randomly choose 10,000 feedback items from our data set (\ie the 
entire 50,378,766 feedback items of our six target services). We
manually read the feedback texts to understand what users typically provide. 
Surprisingly, we find that there is a considerable number (4,450) of the feedback items that are irrelevant to reporting system issues.

\begin{figure}
    \centering 
    \begin{mdframed}
    \begin{enumerate}[leftmargin=1em, label=\roman*), font=\small]
\item \small Dear friend. I have broken up with my girlfriend. My user ID now includes her name. Can you provide a new functionality to the app, so that I can change my ID easily?  I can't live with her name in my user ID. 
\item \small Can you tell me how much money I can transfer per day?
\item \small I strongly suggest adding a yawn emoji!
\item \small I hope there will be a backup function for chat records.
\item \small I'm bored recently. I want to chat with some strangers to kill time.
\item \small Can you hear me? I am very unhappy! My boss is not satisfied with my performance. 
\end{enumerate}
\end{mdframed}
    \caption{Example feedback items irrelevant to system issues.}
    \label{fig:example-irrelevant-feedback}
\end{figure}

We list some irrelevant feedback items as examples, shown in Figure \ref{fig:example-irrelevant-feedback}. 
We can see that the irrelevant feedback items cover a diverse range of topics. Their
percentage in the entire feedback set is comparable to that of the feedback items to report real issues. We summarize three common types of irrelevant feedback items as follows. 

\begin{itemize} 
    \item \textbf{Consultation-related one:} A user may ask some questions about the functionality of the corresponding service in her feedback, \eg items \romannum{1}) and \romannum{2}) in Figure \ref{fig:example-irrelevant-feedback}.
    \item \textbf{Suggestion:} A user may provide her suggestions about the functionality of the corresponding service, \eg items \romannum{3}) and \romannum{4}) in Figure \ref{fig:example-irrelevant-feedback}.
    \item \textbf{Trivial message:} A user may provide a message that is totally irrelevant to the corresponding service, or meaningless words, \eg items \romannum{5}) and \romannum{6}) in Figure \ref{fig:example-irrelevant-feedback}.
\end{itemize}

In contrast, we find that the feedback items that report system issues typically exhibit some common patterns, regardless of the differences in their corresponding services. For example, users usually write ``cannot use ...'' or ``... is down'' when describing the issues they have encountered. 

Note that the above characteristics of the feedback items are obtained by manually analyzing a small set of feedback items (10,000 items). 
It is infeasible to manually examine each feedback item in the entire set. 
To further study the entire feedback set, we have to resort to the automatic method. 

As we witness that the issue-relevant feedback items typically show some common patterns in our manual analysis, we consider that machine learning can be a viable means to automatically identify such feedback items. 

Specifically, we use a binary classification model to determine whether a feedback item reports a system issue.
We first construct a {\em labeled} data set, which consists of two classes of feedback items. One class includes the 4,450 aforementioned feedback items that we have manually marked as irrelevant feedback items, while the other includes 4,450 relevant ones. The latter is obtained by randomly sampling from the items we have marked as relevant feedback. The former and the latter contain the same number of feedback items to meet the class-balance requirement \cite{DBLP:conf/icml/LewisC94}. 
 
We randomly divide the labeled data set into a training set, a validation set, and a test set, which account for 75\%, 15\%, and 15\% of the labeled data set, respectively. Each of these three sets has an equal number of items in either class. This is a common way to partition labeled data sets when training a classification model \cite{DBLP:books/daglib/0082591}

We choose BERT \cite{DBLP:conf/naacl/DevlinCLT19}, a state-of-the-art deep learning-based method, to conduct text embedding. It vectorizes the texts in a feedback item so that we can further apply machine learning-based methods to analyze the texts.   
We then train a TextCNN-based binary classifier (\ie a widely adopted deep learning-based text classification approach \cite{DBLP:conf/emnlp/Kim14}) using our labeled data. It learns from our training dataset how to classify a feedback item as either issue-relevant or irrelevant. In particular, we follow a typical training-validating-testing approach \cite{DBLP:books/daglib/0082591}. We first use the training set to train a set of such classifiers with different parameters (\eg batch size, learning rate) and evaluate their performance on the validation set. Then we choose the classifier with the best performance on the validation set as our resulting classifier. 

Finally, we evaluate the accuracy, precision, and recall of the classifier on the test set and obtain 89.58\%, 91.29\%, and 86.28\%, respectively. Accordingly, we obtain the F1 score, which is 88.71\%. We can see that our resulting classifier has achieved quite a good performance. 
To further prove that the classifier can achieve good performance on real data (\ie a large amount of user feedback), we further calculate the binomial proportion confidence interval \cite{wackerly2014mathematical} of the accuracy. Such a method for estimating accuracy when applying a classifier to real data is commonly used to evaluate a classifier's performance \cite{DBLP:books/daglib/0087929}. We calculate the confidence interval $R$ according to the following equation:
\begin{equation}
\label{eq:confidence_interval}
R = Z \sqrt{\frac{acc(1-acc)}{n}}
\end{equation}
where $acc$ is the classifier accuracy, $n$ is the size of the test set. $Z$ is a constant according to the confidence, which is $1.96$ when we choose 95\%, a conventional level, as the confidence. In this way, we obtain the confidence interval [87.87\%, 91.29\%]. This indicates that there is a 95\% likelihood that the range 87.87\% to 91.29\% covers the ground-truth accuracy of the classifier. 
Hence, we consider the classifier to be accurate enough to automatically analyze whether a feedback item is relevant to system issues. 

We can then apply the classifier to the entire feedback item set. Table \ref{table:feedback} summarizes the results, \ie the total amount of feedback items and issue-relevant ones of our six target services in the entire one-year user feedback set.
We can observe that the issue-relevant feedback items actually do not dominate in the feedback set, in each of the six services. The proportions range from 10.94\% (\work) to 66.54\% (\wechat). The results indicate that we cannot consider all user feedback items to be relevant to system issues in mining the issues from user feedback items (\eg \cite{DBLP:conf/kbse/ZhengLZLZD19,DBLP:conf/issre/GaoWHZZL15}). Feedback items, as they are obtained directly from the general end users without any quality control, may contain too much irrelevant information. 
Therefore, in production scenarios, it is necessary to first automatically filter out irrelevant user feedback items before applying a mining approach to analyze potential issues. In this way, we can avoid being misled by irrelevant items and false-positive issues. Fortunately, our results also show that a deep learning-based approach, \eg one adopted by us, can exhibit good performance in filtering out the irrelevant user feedback items. 

Finally, it is worth noting that the proportion of issue-relevant feedback in different services varies. 
In \work, \info, \pay, and \weread, the issue-relevant feedback items of each service account for less than 30\%. In \wechat and \game, the proportions are over 60\%.
A higher proportion value may indicate a lower service quality. 
For the \game service, this is probably because it hosts many third-party-developed games. Based on the experiences of service developers, the software quality of these games is generally low compared with that of the service platform itself. As a result, users often complain about bad gaming experiences caused by software issues in the games, \eg ``the game often crashes'', ``the game is always lagging'' and ``the game cannot be loaded''. This leads to a high proportion of issue-relevant feedback items.
But for the \wechat service, the reason is different. \wechat requires more steps to submit a feedback item. A user is forced to categorize a feedback item gradually before she can come to the feedback interface. This may discourage those who intend to provide irrelevant information, resulting in a high proportion of issue-relevant feedback items.

\begin{table}[t!]
\footnotesize
\centering
\caption{Total amount of feedback items and issue-relevant feedback items collected from six services in one year.}
\resizebox{0.95\linewidth}{!}{
\begin{tabular}{ l  c  c  c }
 \toprule
 \textbf{Service} &  \textbf{\# Feedback} &  \textbf{\# Issue-relevant} & \textbf{Percentage} \\
 \midrule
 \textsl{WeChat} & 16,405,550 & 10,916,729 & 66.54\% \\
 \textsl{WeChat-Work} & 20,406,139 & 2,232,664 & 10.94\% \\
 \textsl{WeChat-Info} & 7,771,103 & 1,289,212 & 16.59\% \\
 \textsl{WeChat-Pay} & 2,306,460 & 652,204 & 28.28\% \\
 \textsl{WeChat-Game} & 2,897,146 & 1,752,791 & 60.50\% \\
 \textsl{WeChat-Reading} & 592,368 & 174,179 & 29.40\% \\
 \bottomrule
\end{tabular}}
\label{table:feedback}
\end{table}

 \begin{tcolorbox}[colback=gray!10,
                  colframe=black,
                  width=\linewidth,
                  arc=3mm, auto outer arc,
                  boxrule=1pt
                 ]
  \textit{\textbf{Summary of RQ1:} Only 10.94\%$\sim$66.54\% of user feedback items across six services were issue-relevant, revealing the necessity of automated filtering to remove noise before analysis.}

\end{tcolorbox}
\subsection{RQ2: Can the number of feedback items on an issue indicate how severe the issue is?}
\label{finding:rq2}

As shown in the "Issue-relevant" column of Table \ref{table:feedback}, each service can still collect a large volume of issue-relevant feedback items during its runtime. 
Although each of these feedback items may indicate a potential service issue to some extent, it is labor-intensive, if not infeasible, to manually read each item and analyze the potential issue the item describes, given the huge number of such items. 

In addition, the reported issues are not equally important. Some are severe system defects that need to be addressed quickly to avoid a more disastrous influence on the target service. Examples include those reporting the crash of a certain service functionality. 
Some, on the other hand, are trivial, which have little impact on users, \eg that reporting ``one of my friends' new avatar cannot be displayed''. 

Hence, to better exploit the user feedback in issue detection, we should resort to automatic methods for analyzing the feedback items. The key to such methods is generally prioritizing the issues (or, more concretely, the topics in the item contents) provided in the feedback items. In this way, those feedback items that potentially indicate severe issues
can receive prompt attention from the service developers. 

An instant consideration is that we should first group the feedback items according to their content, using natural language processing techniques, for example. Then, based on the notion that more severe issues affect more users, we can rank the groups according to the number of feedback items in each group. 
Such a method seems quite reasonable and is a typical approach adopted in existing work. For example, work in \cite{DBLP:conf/kbse/ZhengLZLZD19} suggests that items can be grouped based on the words they contain, and the groups with a large number of items are the candidate groups that indicate severe issues. 

Hence, we are motivated to examine further whether such a consideration is valid for our target large-scale service system. In other words, we aim to investigate whether real-world severe system issues can accumulate a large number of feedback items. However, studying such a question is quite hard. Feedback items do not always indicate severe issues that deserve further investigation. Determining the severity of the corresponding issue of an item remains an open, challenging task to the community (which we will further discuss in Section \ref{sec: each-item-importance}) \cite{DBLP:conf/kbse/ZhengLZLZD19}. Hence, as we have discussed, in current industry practice, many feedback items in online service systems are simply stored without being further analyzed or labeled with the issues they describe. Our target platform is not an exception. In other words, to answer our research question, \ie whether a real-world severe system issue can always accumulate a large number of feedback items, we cannot just resort to simple calculation on our feedback item set.

Fortunately, we find that the \textit{issue-tracking system} of the target platform can greatly facilitate our analysis of the question.  
In the issue-tracking system, an {\em issue ticket} describes a known issue, and records the information of how the corresponding issue is detected, confirmed, triaged, processed, and fixed, as well as the symptoms, the root cause, the scope of its influence, and its severity description. In addition, if the ticket is initialized by user feedback, it also records the corresponding user feedback items.

We collected all issue tickets (509 in total) with user feedback items during one year.
Since each ticket has been labeled with a severity description, we can accordingly group such tickets into five groups, as shown in Table \ref{table:severity}. 
In this table, we use the level number to indicate the severity of the issues: the lower the number, the more severe the corresponding issue. We labeled the level manually according to its description. 
We intentionally provide such a coarse-grained partition to hide the confidential information (\eg detailed issue and severity description). 
With such a severity-based partition, we can now study whether more severe issues have accumulated more user feedback items. 

\begin{figure}
    \centering
    \includegraphics[width=0.9\linewidth]{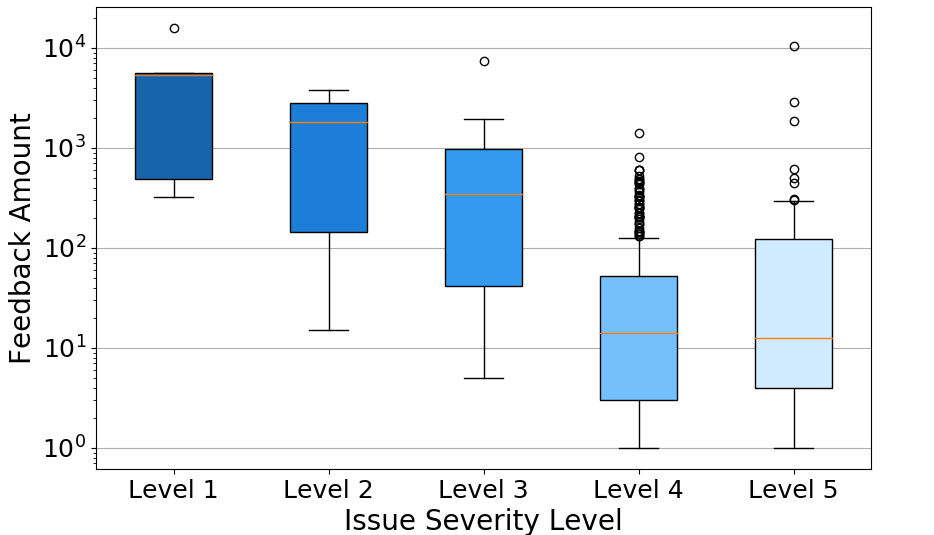}
    \caption{Statistics of feedback items corresponding to different severity levels.}
    \label{fig:severity}
    \vspace{-0.3cm}
\end{figure}

Figure \ref{fig:severity} shows the statistics of the feedback items corresponding to each level of issues with a box plot. Each box is drawn from the 25th percentile to the 75th percentile, with a horizontal line in the box to denote the median. The two horizontal lines outside the box denote the minimum and the maximum value of the data set.
We can see that in such a postmortem analysis, it is generally true that a more severe issue can receive more user feedback items.
The number of feedback items that describe the same issue tends to be a good indicator of how severe the corresponding issue is. 

\begin{table}[t!]
\centering
\caption{The severity distribution of collected issues in one year.}
\resizebox{\linewidth}{!}{
\begin{tabular}{ c  c  c  c  c  c  c }
 \toprule
  &  \textbf{Level 1}  & \textbf{Level 2} & \textbf{Level 3} & \textbf{Level 4} & \textbf{Level 5} & \textbf{Total}\\
 \midrule
 \# Issues & 5 & 10 & 43 & 391 & 60 & 509 \\
 \bottomrule
\end{tabular}}
\label{table:severity}
\end{table}

However, we can also witness many exceptions. Some issues, although receiving only a small number of user feedback items, are still very severe issues that may cause disastrous system failures. 
For example, we can observe that a level-1 issue and a level-2 issue only received 321 and 15 feedback items, respectively. Some issues, on the other hand, although receiving tremendous feedback items, are relatively less-critical issues, \eg we can find that a level-5 (\ie the lowest severity) issue has received up to 10,500 related feedback items.

These results show that, in general, the number of feedback items on a particular issue can, to some extent, indicate the severity of the issue. This shows that prioritizing user feedback issues based on such a feedback item number is a viable means. 
However, since we also witness many exceptional cases, we consider this method for prioritizing feedback items to be still inadequate. 
Especially when a severe issue occurs with only a small number of user feedback items, developers cannot detect it in a timely manner using this method. This may lead to severe consequences (\eg system crash). Hence, it is necessary to work out alternative approaches to cope with such a problem. Next, we will analyze whether it is possible to determine the severity of the issue based on the feedback content. 

\begin{tcolorbox}[colback=gray!10,
                  colframe=black,
                  width=\linewidth,
                  arc=3mm, auto outer arc,
                  boxrule=1pt
                 ]
  \textit{\textbf{Summary of RQ2:} While severe issues generally attract more feedback, exceptions exist where critical issues (e.g., system crashes) were reported by very few users, limiting reliance on volume-based prioritization.
}
\end{tcolorbox}

\subsection{RQ3: Can certain features of a feedback item indicate the severity of its reported issue?}
\label{sec: each-item-importance}

As we have discussed, prioritizing user feedback issues based on the number of corresponding feedback items cannot reveal all system issues. 
Some severe issues may not receive a large amount of user feedback. 
A natural consideration is whether we can analyze a feedback item itself to determine whether it reports a severe issue. Next, we provide our investigation on this research question. 

We first manually read the feedback items (collected in RQ2) that report severe issues. Unfortunately, our experiences show that such items do not exhibit explicit features in their texts. 
In other words, we find it quite difficult to identify the items that report severe issues by comprehending their texts. We confirm our considerations by interviewing the supporting engineers of our target services. Their experiences also show that the severity of a reported issue is typically related to the specific service. 
It generally requires domain-specific knowledge to manually inspect whether a reported issue is severe, which may also involve inspecting code or logs. 

In this regard, we are motivated to investigate whether other features in 
a feedback item may help determine severe issues. Specifically,
we conduct an empirical study on three features. 
The first is the {\em sentiment} of the feedback item: A user may tend to express negative sentiments when reporting a severe issue. 
The second is the {\em text length} of the item: When a user reports a severe issue, she may provide more information, resulting in a longer feedback text. 
The third one is the {\em historical behaviors} of the user: The issues reported by the users who have reported severe issues before are more likely to be severe ones, compared with those 
reported by other users. 

To study whether the text-based features of a feedback item (\ie its sentiment 
and text length) are good indicators of severity, we resort to statistical methods. 
In particular, we first consider a hypothesis that assumes a feature (\eg sentiment) is {\em not} an indicator of severe issues. We then conduct hypothesis testing to examine whether the hypothesis should be accepted or rejected.
What follows provides the details of our study.

\subsubsection{\textbf{Sentiment}}
We first conduct a sentiment analysis on the issue-relevant feedback texts. 
We apply SnowNLP, a widely adopted sentiment analysis tool \cite{github:snownlp}, to calculate a sentiment score (ranging from 0 to 1) for each feedback text. 
We take a common strategy to partition the score range into three sentiment intervals: negative sentiment interval (less than 0.2), neutral one (between 0.2 and 0.8), and positive one (larger than 0.8) \cite{DBLP:journals/coling/TaboadaBTVS11}. In this way, we can divide the feedback items for each of our six target services 
into three groups, the negative group (Neg.), the neutral one (Neu.), and the positive one (Pos.). 
We randomly sample 240 feedback items from each group, and thus obtain 720 items for each target service.

We then evaluate whether the items in a negative group are more likely to indicate severe issues by conducting a \textit{Z-test} on the data. 
The Z-test is a classic proportion hypothesis testing method widely used to determine whether the proportions of two groups are significantly different based on sampling \cite{wackerly2014mathematical}. 
The key to the Z-test is to obtain the Z value that determines the significance of the difference, as follows.
\begin{equation}\label{eq:z-test}
\begin{aligned}
    p & =\frac{p_1n_1+p_2n_2}{n_1+n_2} \\
    Z & =\frac{p_1-p_2}{\sqrt{p(1-p)(\frac{1}{n_1}+\frac{1}{n_2})}}
\end{aligned}
\end{equation}
where $n_1$ and $n_2$ are the number of samples in two groups, $p_1$ and $p_2$ are the proportion of the feedback items that potentially indicate severe issues. 

To conduct a Z-test, we should first label the items that potentially indicate severe issues (\ie obtain $p_1$ and $p_2$ in Equation (\ref{eq:z-test}). We resort to manual inspection of the 720 items for each target service. Based on our interview with support engineers, we find that a feedback item that potentially reports severe issues should 1) describe seemingly severe issues (\eg crash/not-response), and 2) clearly describe the occurring contexts (\ie in what scenario
such an issue occurs).  We conduct our manual inspection based on such criteria.

\begin{table}[t!]
\footnotesize
\centering
\caption{Proportions of feedback items indicating severe issues in different sentiment groups and the Z-test results.}
\label{table:sentiment}
\resizebox{\linewidth}{!}{
\begin{tabular}{ l  c  c  c  c  c }
 \toprule
 \multirow{2}{*}{\textbf{Service}} & \multicolumn{3}{c}{\textbf{Sentiment}} & \multicolumn{2}{c}{\textbf{Z value}} \\
 \cmidrule{2-6}
 & Neg. & Neu. & Pos. & Neg.-Neu. & Neg.-Pos. \\
 \midrule
 \textsl{WeChat} & 24.2\% & 22.5\% & 19.6\% & 0.432 & 1.215  \\
 \textsl{WeChat-Work} & 31.7\% & 24.6\% & 32.5\% & 1.308  & -0.196  \\
 \textsl{WeChat-Info} & 24.2\% & 20.8\% & 28.3\% & 0.874  & -1.037  \\
 \textsl{WeChat-Pay} & 19.6\% & 12.5\% & 14.7\% & \textbf{2.114}  & 1.388  \\
 \textsl{WeChat-Game} & 39.2\% & 28.8\% & 17.5\% & \textbf{2.410}  & \textbf{5.267}  \\
 \textsl{WeChat-Reading} & 25.8\% & 23.3\% & 20.0\% & 0.636  & 1.520  \\
 \bottomrule
\end{tabular}}
\\[3pt]
* Neg.: Negative group, Neu.: Neutral group, Pos.: Positive group
\end{table}

Table \ref{table:sentiment} shows the percentage of the feedback items that potentially 
report severe issues. We can instantly observe that in the six services, the proportions of 
such feedback items in the three sentiment groups do not show much difference.

Table \ref{table:sentiment} further shows the results of the Z-test, in which we study
whether the proportions of the feedback items that potentially indicate severe issues are similar between the negative group and either of the other two groups (the neutral one and the positive one).
We select 0.05 as the significance level $\alpha$ of the test, which is a commonly-adopted value \cite{wackerly2014mathematical}. In such a setting, the hypothesis (that sentiment is irrelevant to whether a feedback item potentially indicates a severe issue) can be rejected if the Z value is larger than 1.65. 

We can see that among the six services, only in the \game service, negative sentiment can be a factor that potentially indicates severe issues. This is reasonable, the \game service hosts many third-party games. The quality of these codes is not the same as that of the self-developed codes.
As a result, low-quality games cause many users to report issues and complain about bad game experiences. Such user feedback presents negative sentiments.
However, in general, sentiment is not a good criterion for identifying feedback items that potentially report severe issues.

\subsubsection{\textbf{Text Length}}
We now study another text-based feature, the text length of the feedback items, with a similar method. We obtain the text length of feedback items for each service. We divide the items into three groups again. 
The short-text group (S.) contains items with less than 20 Chinese characters.
The medium-text group (M.) contains those with between 20 and 50 characters. The long-text group (L.) includes those with more than 50 characters. 
Again, we randomly sampled 240 feedback items from each group and obtained 720 items for each target service.

We also conduct manual inspections of the items and label the items that potentially indicate severe issues, similarly to what we have done in analyzing the sentiment feature. 
We thus obtain the proportion of such items in each group, as shown in Table \ref{table:length}.

\begin{table}[t!]
\centering
\caption{Proportions of feedback items indicating severe issues in different text-length groups and the Z-test results.}
\label{table:length}
\resizebox{0.9\linewidth}{!}{
\begin{tabular}{ l  c  c  c  c  c}
 \toprule
 \multirow{2}{*}{\textbf{Service}} & \multicolumn{3}{c}{\textbf{Text-length}} &  \multicolumn{2}{c}{\textbf{Z value}} \\
 \cmidrule{2-6}
 & S. & M. & L. & L.-M. & L.-S. \\
 \midrule
 \textsl{WeChat} & 20.6\% & 24.0\% & 26.8\% & 0.615  & 1.611  \\
 \textsl{WeChat-Work} & 25.2\% & 45.1\% & 37.1\% & -1.388  & \textbf{2.755}  \\
 \textsl{WeChat-Info} & 22.8\% & 27.8\% & 28.8\% & 0.083  & 1.319  \\
 \textsl{WeChat-Pay} & 13.0\% & 25.0\% & 18.9\% & -1.207  & \textbf{1.727}  \\
 \textsl{WeChat-Game} & 29.4\% & 25.8\% & 31.4\% & 1.155  & 0.462  \\
 \textsl{WeChat-Reading} & 21.3\% & 21.5\% & 27.5\% & 1.376  & 1.543  \\
 \bottomrule
\end{tabular}}
\\[3pt]
* S.: Short-text group,  M.: Medium-text group, L.: Long-text group

\end{table}

Again, we resort to the Z-test. The results are also shown in Table \ref{table:length}, which has a setting similar to our study on the sentiment feature.
It shows that the hypothesis (that text length is irrelevant to whether a feedback item potentially indicates a severe issue) cannot be rejected in general.
In other words, text length is not a good criterion to identify the feedback items that potentially 
report severe issues. Users are allowed to report long texts, even for less critical issues. Here is an example of such feedback items for the \wechat service:

\begin{itemize}
    \item[] \textit{"Dear stuff, I receive no anonymous greeting for days. Please help me find out what's going on. I really need new friends. Thank you for solving my problems."}
\end{itemize}

\subsubsection{\textbf{Historical Behaviors}}
Finally, we study whether users' historical behaviors in providing feedback can be a good indicator of severe issues. As we intend to obtain our results with more cases, we focus on the \wechat service since this service has far more users and feedback items. 
It contains more severe issues, and meanwhile, more users have reported more than one issue. 

We first analyze the 159 feedback items of \wechat that potentially report severe issues, which we have manually labeled in studying the sentiment feature. These items are provided 
by 159 users. 
We examine the historical behaviors of these users. We find that 79 of them only provided one feedback within a year. Then, for the remaining 80 users, we can manually inspect their multiple feedback items. We find that 9 of them have provided feedback on other severe issues within a year. The average number of such feedback items is close to three. 
This result shows that user behaviors can, to some extent, indicate the severity of the issue a user reports. A proper persona model of users is, hence, promising in evaluating the severity of the issues they report.

However, it is worth noting that the number of users who provide frequent feedback is still rare. In addition, the issues they report are merely a small proportion of all reported severe issues. As a result, we cannot rely only on user behavior analysis to check whether an item reports a severe issue. 

\begin{tcolorbox}[colback=gray!10,
                  colframe=black,
                  width=\linewidth,
                  arc=3mm, auto outer arc,
                  boxrule=1pt
                 ]
  \textit{\textbf{Summary of RQ3:} Text-based features (sentiment, text length) showed negligible correlation with issue severity, but historical user behavior (e.g., prior severe-issue reports) can offer some predictive value.
}
\end{tcolorbox}

\subsection{RQ4: Will the topics of user feedback change significantly over time?}

As we have shown in our previous discussions, analyzing the texts in the feedback items is necessary, for example, to examine whether an item is issue-relevant or to group the items according to the issues they describe. 
With the recent development of machine learning techniques, especially deep learning techniques, text analysis is typically conducted with machine learning-based natural language processing approaches~\cite{DBLP:conf/icml/CollobertW08}. 

One key basis of these learning-based approaches is that the characteristics of the texts remain stable within the time intervals of interest.  
In other words, one can learn the characteristics of the texts in one past interval, form a machine learning model (\eg a classifier), and then apply the model to analyze the texts in the future interval. 
If the text characteristics of the items vary over time, such learning-based approaches may not produce good performance in analyzing future, unknown texts. 
Hence, we are motivated to examine whether the feedback items are stable in their topic distribution, with the tremendous number of feedback items we collect. 
 
As we aim to study whether user feedback changes over time, we intentionally 
focus on the items in different intervals that are more likely to be different. 
To this end, for each service, we compare the feedback items along the service upgrade track. 
We divide the intervals based on service code versions because new features are typically introduced in new versions, and user feedback topics may change more frequently compared to those for the same version. 

We select eight versions {\em evenly} from tens of versions released during one year for each service, so that the differences between the release dates of any two versions are over one month. Our aim is to maximize the version differences among these eight versions. 
For each version, we focus on issue-relevant feedback items within one week after the version release. We focus on the user feedback in this period since we consider that such feedback is more likely on the issues in a new version. 
In each service, the average amount of collected feedback items in each version is about 10,000. 
Then we can compare the user feedback items in these eight groups for each service.

Since the feedback items of the two groups are, by nature, two sets of random data, their difference can be measured by the difference of their distributions. In this regard, we again vectorize the texts of each feedback item into a 768-dimensional vector using BERT, similarly to what we have done in Section \ref{finding:rq1}. In this way, we model the 
topic of each item as a point in a 768-dimensional space. We can then use \textit{Wasserstein Distance}~\cite{article:wasserstein} to measure the similarity of the feedback items in two different groups. 

The Wasserstein Distance is a widely accepted metric for measuring the similarity between any two distributions. It essentially models the distance between two distributions as the cost of changing one distribution to another. 
Note that other metrics, \eg  Kullback-Leibler Divergence \cite{article:kldivergence} and Jensen-Shannon Divergence \cite{DBLP:journals/tit/Lin91} may also be viable candidates. A Wasserstein Distance is in the interval [$0$, $+\infty$], where the smaller the value, the more similar the two distributions.
We employ the \textit{Sinkhorn Algorithm} \cite{DBLP:conf/nips/Cuturi13}, a fast algorithm widely used to calculate the approximate Wasserstein Distance. 

To obtain a baseline for comparison purposes, for each of the eight groups, we randomly divide the feedback items into two parts.
We calculate the Wasserstein Distance of the two parts in each of the 8 groups. In this way, we obtain an interval that contains the eight Wasserstein Distance values, which we take as a baseline. 
The underlying consideration is that the items in these two parts, since they are collected in the same interval, have the same distribution. Hence, their  Wasserstein Distance values can serve as a baseline. 

\begin{figure}
    \centering
    \includegraphics[width=\linewidth]{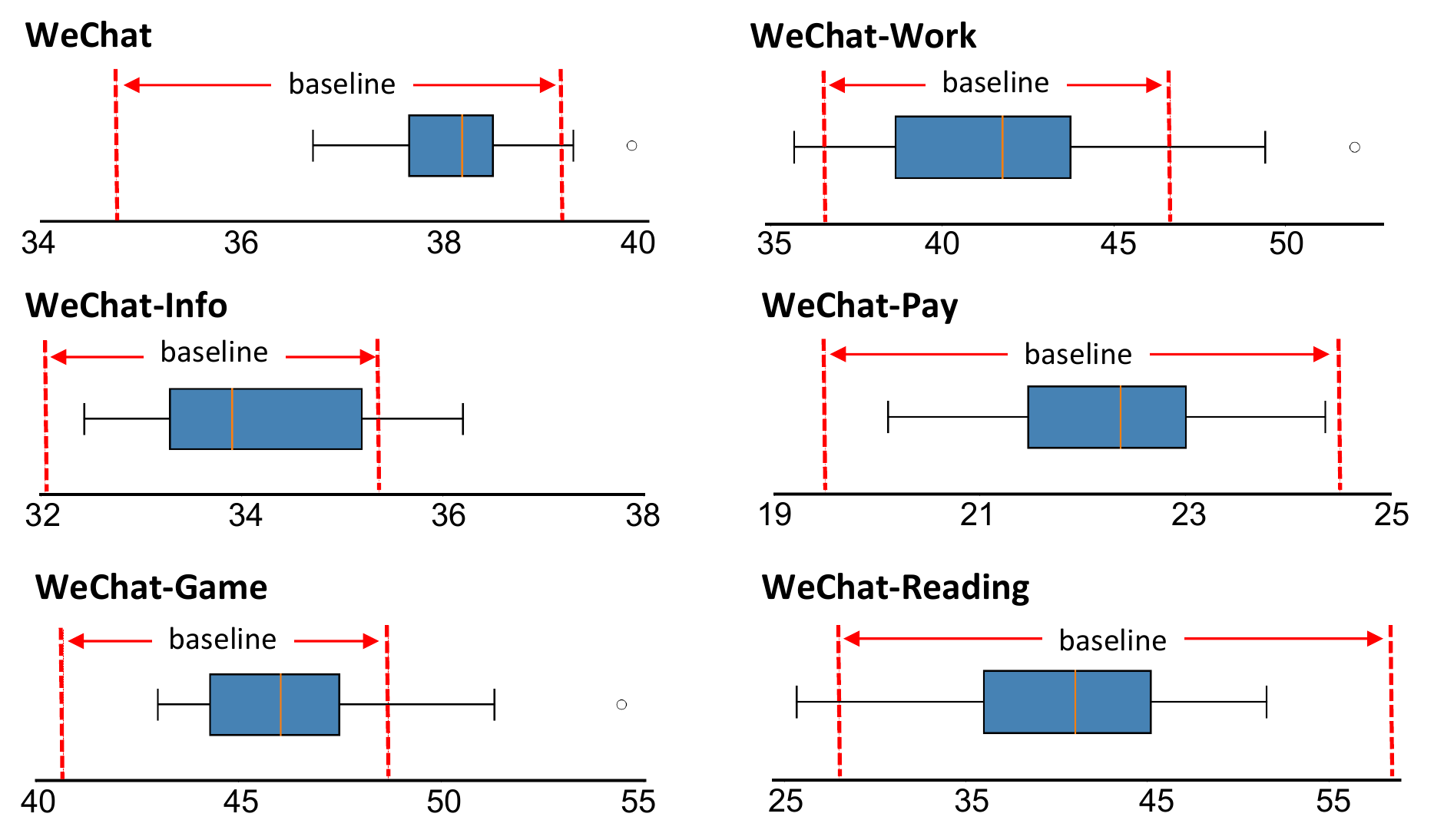}
    \caption{Wasserstein Distance results of the six target services.}
    \label{fig:wd}
\end{figure}
   
Thus, for each service, we obtain the Wasserstein Distance between each two of the eight groups, for a total of 28 distance values. 
Figure \ref{fig:wd} illustrates the Wasserstein Distance results of the six target services.
We can observe that in each service, the Wasserstein Distance values of the feedback distributions among different groups are generally close to the baseline. 
This means that the feedback distributions of any two groups are similar. 
In other words, the feedback topics of these six services are generally stable over time. These results suggest that it is reasonable to apply machine learning-based approaches for user feedback analysis.

\begin{tcolorbox}[colback=gray!10,
                  colframe=black,
                  width=\linewidth,
                  arc=3mm, auto outer arc,
                  boxrule=1pt
                 ]
  \textit{\textbf{Summary of RQ4:} Feedback topic distributions remained stable across service versions and time intervals (Wasserstein Distance $\approx$ baseline), validating the feasibility of machine learning for longitudinal analysis.
}
\end{tcolorbox}

\section{Further Discussions}\label{sec:discussion}

\subsection{Lessons Learned}
As discussed, we conducted an empirical study on all the user feedback items collected over a one-year period for six real-world services in a one-billion-user online service system. We first investigate what users will provide in their feedback items, with the aim of examining whether they are reporting real system issues. 
We find that there is a lot of irrelevant information in the feedback. Only a small proportion (which may be lower than $30\%$) of the feedback items are issue-relevant. In this regard, a filtering mechanism (\eg a classifier to determine whether a feedback item is issue-relevant) is essential for feedback-based issue detection.

We then examine how to conduct issue detection in a large volume of issue-relevant feedback items. 
To this end, we investigate whether the severity of the reported issue is related to the number of its corresponding feedback items. Such a relation is a general basis for many issue-mining approaches \cite{DBLP:conf/icse/GaoZLK18, DBLP:conf/icse/GaoZD0ZLK19}. We find that it is generally true that a severe issue can receive more feedback items. However, we also observe exceptions. 
This means that if we rely only on the feedback item numbers, we may neglect some severe issues reported in some user feedback items. 

Hence, we further investigate whether a feedback item itself can provide severity information on the issue. 
We, in particular, study the influences on issue severity of three features of an item, \ie sentiment, text length, and historical user behaviors. Unfortunately, we find that the text-based features are, in general, not good criteria for identifying potential severe issues. However, a proper model of the user can, to some extent, help determine whether the issue she reports is severe.  

Finally, as we find that text analysis (\ie typically natural language processing with machine learning) is critical to issue analysis in user feedback items, we further examine whether such a learning-based approach is reasonable in analyzing real-world user feedback. 
To this end, we investigate whether the experiences learned from the past still apply to the future. Specifically, we study the distributions of the feedback topics in different time intervals. To justify the learning-based approach, we examine whether such distributions are similar. 
Our results confirm that in real-world user feedback, the topics of feedback are generally stable. Users tend to report issues using similar language, which suggests that a learning-based approach is a viable way to analyze issues from user feedback.

\subsection{Threats to Validity}

We now discuss possible threats to our study and the methods we use to address them. 
One possible threat is that our results may not apply to other online service systems.
We mitigate this threat by choosing WeChat, a large-scale social media platform with a tremendous user base, as our target system. 
We consider that the user behaviors (\eg how users describe issues in their feedback) are similar among different cloud systems. We examine the feedback provided by a huge number of users, which allows us to better understand general user behaviors. 
Moreover, we choose six services with diverse functionalities (\eg work and entertainment). Thus, we can avoid misleading results caused by the service specifics.

Another threat is that we may be misled by our source data (\ie the feedback items) if they exhibit a certain feature in some particular interval \cite{DBLP:conf/msr/MartinHJSZ15}. 
For example, our result may be obtained when our target services undergo upgrades. 
We address this threat by considering the entire set of user feedback data collected in one year. 
For each service, the collected data spans multiple service versions. We can thus minimize the risk of being misled by potentially biased data. 

Finally, some of our findings are based on manual inspections. For example, we label whether a feedback item reports a severe issue by manual inspection. Lack of domain knowledge in labeling the data may, to some extent, influence the results. We alleviate such influence with two steps. We first conduct interviews with expert service support engineers and learn how to label severe issues. Second, we ask them to confirm our manual inspection results. In this way, we can best guarantee the correctness of the results.

\subsection{Design Implications of Feedback-based Issue Analysis}

In our investigation of RQ1, we find that most services have a small 
issue-relevant feedback proportion. In contrast, the proportion for the \wechat services is much larger. 
\wechat requires users to first categorize the reported issue before submitting it. 
It has long been suggested that design factors (\eg app affordances and information visualization) affect user perception and thus change user behaviors \cite{DBLP:conf/chi/SrinivasanBLD18, DBLP:conf/chi/GrayKBHT18}. 
We can see that the feedback mechanism in \wechat has a positive influence on the quality of feedback. 
This inspires us that a well-designed feedback interface may effectively increase the proportion of issue-relevant feedback. 
Irrelevant feedback is misleading and hence should be discouraged. 
We call for more future studies on the factors affecting the quality of user feedback, and new feedback collection mechanisms that encourage more useful feedback.

In addition, our results reveal that the topic distributions of user feedback are stable over time. Hence, it is a promising way to adopt machine learning techniques to process user feedback. Specifically, we can train a machine learning model using historical feedback and then apply it to newly arriving feedback. 
For instance, we can train a classifier based on existing feedback that is issue-relevant or irrelevant, which can then be used to filter future feedback items. Moreover, due to the huge volume of feedback items, it is also a feasible way to train more complex machine learning models (\eg transformer-based model \cite{DBLP:journals/corr/abs-2102-10772}) to conduct more sophisticated analysis (\eg with semantic analysis \cite{DBLP:conf/ijcai/WangWZY17} or with deep language model \cite{DBLP:conf/naacl/DevlinCLT19}) on feedback items. We suggest that these tasks are important future directions that deserve more research efforts.

Finally, we find it quite common that severe issues exist that cannot be easily detected with only user feedback. 
In particular, prioritizing issues based on the number of feedback items may neglect severe issues with a small number of feedback items.
On the other hand, the characteristics of one single feedback item, \eg text features like sentiment and text length, are also not good criteria in identifying severe issues.
Although historical user behaviors can help indicate severe issues, this method only covers a small proportion of users and severe issues. 
These phenomena show that corner cases always exist when trying to identify severe issues based solely on user feedback. 
In other words, feedback is not what we can entirely rely on. 
Fortunately, user feedback is not the only resource for identifying issues. Large-scale service systems typically adopt a large number of system monitors to examine specifically tailored key performance indicators (KPIs). Automated, real-time issue detection can be conducted based on these KPIs \cite{DBLP:journals/tpds/MiWZLC13}. Although existing work shows that system monitors are also not enough to detect issues \cite{DBLP:conf/kbse/ZhengLZLZD19}, jointly considering user feedback and the KPI values may be a viable means to detect more system issues. For instance, recently, LinkCM \cite{DBLP:conf/sigsoft/GuWWZLKZYSXQLLZ20} has been proposed to link user-reported issues to these KPIs, aiming to achieve better issue detection. We consider this to be also a critical research direction.
 
\section{Related Work}\label{sec:related}
User feedback serves as a communication bridge between developers and users. It has long been suggested that user feedback can be of great value in supporting software development and maintenance \cite{DBLP:conf/icsm/PalombaVBOPPL15, DBLP:conf/kdd/FuLLFHS13, stradowski2023industrial, he2025practicability, tan2023stre, chen2024smart}. 

A body of existing work focuses on analyzing user reviews from app markets (\eg Google Play and Apple's App Store) to help app development in various aspects \cite{DBLP:journals/infsof/FinkelsteinHJMS17}. Iacob \etal~\cite{DBLP:conf/bcshci/IacobVH13} identify the recurring issues via manually labeling 3,278 user reviews of 161 apps. Since analyzing a large number of user reviews manually is labor-intensive, many previous studies resort to automated methods to extract app features from user reviews \cite{DBLP:conf/kbse/VuNPN15, DBLP:conf/kbse/VuPNN16, DBLP:conf/issre/ManGLJ16}. For instance, Vu \etal~\cite{DBLP:conf/kbse/VuPNN16} propose PUMA, which extracts user opinions from app reviews via a phrase-based clustering approach. Man \etal \cite{DBLP:conf/issre/ManGLJ16} collect descriptive words for specific app features through a text vectorization-based approach. 

In addition to these approaches that focus on extracting app features, another line of work focuses on summarizing user requirements from user feedback so as to assist developers in software maintenance \cite{DBLP:conf/icse/VillarroelBROP16, DBLP:conf/www/LuizVAMSCGR18,DBLP:conf/sigsoft/SorboPASVCG16,DBLP:conf/re/MaalejN15}. For example, Di Sorbo \etal \cite{DBLP:conf/sigsoft/SorboPASVCG16} propose SURF, a system that employs machine learning techniques to conduct topic classification and summarize user requirements from user reviews. Luiz \etal~\cite{DBLP:conf/www/LuizVAMSCGR18} propose a framework to automatically extract features and analyze review sentiment, allowing developers to improve the app in a user-centric manner. There are also other studies aiming at identifying app issues \cite{DBLP:conf/issre/GaoWHZZL15, DBLP:conf/icse/GaoZLK18}, detecting device- or platform-related issues \cite{DBLP:conf/issre/ManGLJ16, DBLP:conf/icse/LuLLXMH0F16}, and supporting the evolution of apps \cite{DBLP:journals/jss/PalombaVBOPPL18}. However, all these studies focus on user reviews of apps with postmortem methods. Compared to user feedback from large-scale online systems, app user reviews are on a smaller scale. The proposed approaches may not be directly applicable to online issue detection \cite{DBLP:conf/kbse/ZhengLZLZD19}. 

For feedback-based issue detection, a body of work aims to solve the challenges caused by huge system scale and massive data \cite{DBLP:conf/icdm/LimLZFTLDZ14}. Most work in this field designs automated mechanisms to detect runtime incidents based on issue reports \cite{DBLP:conf/icse/LinLZZ16, DBLP:conf/icse/WuWCZ18, DBLP:conf/sigsoft/GuLQQL0LDCWZCZ20} or system logs \cite{DBLP:conf/icse/LinZLZC16, DBLP:conf/sigsoft/HeLLZLZ18}.
Meanwhile, some research has taken advantage of massive user feedback in online services to assist in real-time software issue identification \cite{DBLP:conf/kbse/ZhengLZLZD19, DBLP:conf/icse/GaoZD0ZLK19}. Specifically, Gao \etal propose DIVER \cite{DBLP:conf/icse/GaoZD0ZLK19}, a system that extracts emerging word collocations to detect emerging issues of popular apps. Similar to DIVER, iFeedback \cite{DBLP:conf/kbse/ZhengLZLZD19} automatically extracts word combinations from massive user feedback as key performance indicators and then conducts real-time issue detection based on the indicators. 
However, there is still a lack of comprehensive understanding of real-world user feedback in large-scale online services, which motivates this work.

\section{Conclusion}\label{sec:conclusion}

In this paper, we conduct an empirical study on user feedback from a large-scale online service system that serves billions of users, focusing on the characteristics of feedback for issue detection. By answering four specifically tailored research questions, we find that a large proportion of feedback in real systems does not report issues. Hence, feedback filtering is critical when analyzing feedback. We also find that it is infeasible to identify all severe issues simply based on the number of feedback items or certain features of a feedback item.
Moreover, our study shows that the topic of user feedback is stable over time, which suggests that it is reasonable to adopt machine learning-based approaches to analyze user feedback. These findings provide design implications for feedback-based issue analysis, including designing feedback collection mechanisms, feedback filtering, and jointly considering system KPIs with user feedback for issue detection. This study can serve as an empirical foundation for designing and implementing feedback-based issue detection in large-scale online service systems.

\section*{Acknowledgment}
The work described in this paper was supported by the Research Grants Council of the Hong Kong Special Administrative Region, China (No. CUHK 14206921 of the General Research Fund), and RGC Grant for Theme-based Research Scheme Project (RGC Ref. No. T43-513/23-N).


\balance
\bibliographystyle{IEEEtran}
\bibliography{reference}

\begin{thebibliography}{10}
\providecommand{\url}[1]{#1}
\csname url@samestyle\endcsname
\providecommand{\newblock}{\relax}
\providecommand{\bibinfo}[2]{#2}
\providecommand{\BIBentrySTDinterwordspacing}{\spaceskip=0pt\relax}
\providecommand{\BIBentryALTinterwordstretchfactor}{4}
\providecommand{\BIBentryALTinterwordspacing}{\spaceskip=\fontdimen2\font plus
\BIBentryALTinterwordstretchfactor\fontdimen3\font minus
  \fontdimen4\font\relax}
\providecommand{\BIBforeignlanguage}[2]{{%
\expandafter\ifx\csname l@#1\endcsname\relax
\typeout{** WARNING: IEEEtran.bst: No hyphenation pattern has been}%
\typeout{** loaded for the language `#1'. Using the pattern for}%
\typeout{** the default language instead.}%
\else
\language=\csname l@#1\endcsname
\fi
#2}}
\providecommand{\BIBdecl}{\relax}
\BIBdecl

\bibitem{DBLP:conf/re/PaganoM13}
D.~Pagano and W.~Maalej, ``User feedback in the appstore: An empirical study,''
  in \emph{21st {IEEE} International Requirements Engineering Conference, {RE}
  2013, Rio de Janeiro-RJ, Brazil, July 15-19, 2013}.\hskip 1em plus 0.5em
  minus 0.4em\relax {IEEE} Computer Society, 2013, pp. 125--134.

\bibitem{DBLP:conf/icse/ChenLHXZ14}
N.~Chen, J.~Lin, S.~C.~H. Hoi, X.~Xiao, and B.~Zhang, ``Ar-miner: mining
  informative reviews for developers from mobile app marketplace,'' in
  \emph{36th International Conference on Software Engineering, {ICSE} '14,
  Hyderabad, India - May 31 - June 07, 2014}.\hskip 1em plus 0.5em minus
  0.4em\relax {ACM}, 2014, pp. 767--778.

\bibitem{DBLP:conf/wsdm/NguyenLT15}
T.~Nguyen, H.~W. Lauw, and P.~Tsaparas, ``Review synthesis for micro-review
  summarization,'' in \emph{Proceedings of the Eighth {ACM} International
  Conference on Web Search and Data Mining, {WSDM} 2015, Shanghai, China,
  February 2-6, 2015}.\hskip 1em plus 0.5em minus 0.4em\relax {ACM}, 2015, pp.
  169--178.

\bibitem{DBLP:conf/icse/VillarroelBROP16}
L.~Villarroel, G.~Bavota, B.~Russo, R.~Oliveto, and M.~D. Penta, ``Release
  planning of mobile apps based on user reviews,'' in \emph{Proceedings of the
  38th International Conference on Software Engineering, {ICSE} 2016, Austin,
  TX, USA, May 14-22, 2016}.\hskip 1em plus 0.5em minus 0.4em\relax {ACM},
  2016, pp. 14--24.

\bibitem{lyu1996handbook}
M.~R. Lyu \emph{et~al.}, \emph{Handbook of Software Reliability
  Engineering}.\hskip 1em plus 0.5em minus 0.4em\relax IEEE computer society
  press CA, 1996, vol. 222.

\bibitem{DBLP:conf/kbse/ZhengLZLZD19}
W.~Zheng, H.~Lu, Y.~Zhou, J.~Liang, H.~Zheng, and Y.~Deng, ``ifeedback:
  Exploiting user feedback for real-time issue detection in large-scale online
  service systems,'' in \emph{34th {IEEE/ACM} International Conference on
  Automated Software Engineering, {ASE} 2019, San Diego, CA, USA, November
  11-15, 2019}.\hskip 1em plus 0.5em minus 0.4em\relax {IEEE}, 2019, pp.
  352--363.

\bibitem{DBLP:conf/icse/PaganoB13}
D.~Pagano and B.~Br{\"{u}}gge, ``User involvement in software evolution
  practice: a case study,'' in \emph{35th International Conference on Software
  Engineering, {ICSE} '13, San Francisco, CA, USA, May 18-26, 2013}.\hskip 1em
  plus 0.5em minus 0.4em\relax {IEEE} Computer Society, 2013, pp. 953--962.

\bibitem{DBLP:conf/issre/GaoWHZZL15}
C.~Gao, B.~Wang, P.~He, J.~Zhu, Y.~Zhou, and M.~R. Lyu, ``{PAID:} prioritizing
  app issues for developers by tracking user reviews over versions,'' in
  \emph{26th {IEEE} International Symposium on Software Reliability
  Engineering, {ISSRE} 2015, Gaithersbury, MD, USA, November 2-5, 2015}.\hskip
  1em plus 0.5em minus 0.4em\relax {IEEE} Computer Society, 2015, pp. 35--45.

\bibitem{DBLP:conf/www/LuizVAMSCGR18}
W.~Luiz, F.~Viegas, R.~O. de~Alencar, F.~Mour{\~{a}}o, T.~Salles, D.~B.~F.
  Carvalho, M.~A. Gon{\c{c}}alves, and L.~C. da~Rocha, ``A feature-oriented
  sentiment rating for mobile app reviews,'' in \emph{Proceedings of the 2018
  World Wide Web Conference on World Wide Web, {WWW} 2018, Lyon, France, April
  23-27, 2018}.\hskip 1em plus 0.5em minus 0.4em\relax {ACM}, 2018, pp.
  1909--1918.

\bibitem{DBLP:journals/software/KhalidSNH15}
H.~Khalid, E.~Shihab, M.~Nagappan, and A.~E. Hassan, ``What do mobile app users
  complain about?'' \emph{{IEEE} Softw.}, vol.~32, no.~3, pp. 70--77, 2015.

\bibitem{DBLP:conf/icse/GaoZLK18}
C.~Gao, J.~Zeng, M.~R. Lyu, and I.~King, ``Online app review analysis for
  identifying emerging issues,'' in \emph{Proceedings of the 40th International
  Conference on Software Engineering, {ICSE} 2018, Gothenburg, Sweden, May 27 -
  June 03, 2018}.\hskip 1em plus 0.5em minus 0.4em\relax {ACM}, 2018, pp.
  48--58.

\bibitem{DBLP:journals/jss/PalombaVBOPPL18}
F.~Palomba, M.~L. V{\'{a}}squez, G.~Bavota, R.~Oliveto, M.~D. Penta,
  D.~Poshyvanyk, and A.~D. Lucia, ``Crowdsourcing user reviews to support the
  evolution of mobile apps,'' \emph{J. Syst. Softw.}, vol. 137, pp. 143--162,
  2018.

\bibitem{DBLP:conf/icse/GaoZD0ZLK19}
C.~Gao, W.~Zheng, Y.~Deng, D.~Lo, J.~Zeng, M.~R. Lyu, and I.~King, ``Emerging
  app issue identification from user feedback: experience on wechat,'' in
  \emph{Proceedings of the 41st International Conference on Software
  Engineering: Software Engineering in Practice, {ICSE} {(SEIP)} 2019,
  Montreal, QC, Canada, May 25-31, 2019}.\hskip 1em plus 0.5em minus
  0.4em\relax {IEEE} / {ACM}, 2019, pp. 279--288.

\bibitem{DBLP:books/daglib/0087929}
T.~M. Mitchell, \emph{Machine learning, International Edition}, ser.
  McGraw-Hill Series in Computer Science.\hskip 1em plus 0.5em minus
  0.4em\relax McGraw-Hill, 1997.

\bibitem{wechat}
WeChat, ``Wechat,'' \url{https://www.wechat.com/en/}, 2025.

\bibitem{DBLP:conf/icml/LewisC94}
D.~D. Lewis and J.~Catlett, ``Heterogeneous uncertainty sampling for supervised
  learning,'' in \emph{Machine Learning, Proceedings of the Eleventh
  International Conference, Rutgers University, New Brunswick, NJ, USA, July
  10-13, 1994}.\hskip 1em plus 0.5em minus 0.4em\relax Morgan Kaufmann, 1994,
  pp. 148--156.

\bibitem{DBLP:books/daglib/0082591}
B.~D. Ripley, \emph{Pattern Recognition and Neural Networks}.\hskip 1em plus
  0.5em minus 0.4em\relax Cambridge University Press, 1996.

\bibitem{DBLP:conf/naacl/DevlinCLT19}
J.~Devlin, M.~Chang, K.~Lee, and K.~Toutanova, ``{BERT:} pre-training of deep
  bidirectional transformers for language understanding,'' in \emph{Proceedings
  of the 2019 Conference of the North American Chapter of the Association for
  Computational Linguistics: Human Language Technologies, {NAACL-HLT} 2019,
  Minneapolis, MN, USA, June 2-7, 2019, Volume 1 (Long and Short
  Papers)}.\hskip 1em plus 0.5em minus 0.4em\relax Association for
  Computational Linguistics, 2019, pp. 4171--4186.

\bibitem{DBLP:conf/emnlp/Kim14}
Y.~Kim, ``Convolutional neural networks for sentence classification,'' in
  \emph{Proceedings of the 2014 Conference on Empirical Methods in Natural
  Language Processing, {EMNLP} 2014, October 25-29, 2014, Doha, Qatar, {A}
  meeting of SIGDAT, a Special Interest Group of the {ACL}}.\hskip 1em plus
  0.5em minus 0.4em\relax {ACL}, 2014, pp. 1746--1751.

\bibitem{wackerly2014mathematical}
D.~Wackerly, W.~Mendenhall, and R.~L. Scheaffer, \emph{Mathematical statistics
  with applications}.\hskip 1em plus 0.5em minus 0.4em\relax Cengage Learning,
  2014.

\bibitem{github:snownlp}
\BIBentryALTinterwordspacing
isnowfy. (2017) Snownlp: Simplified chinese text processing. [Online].
  Available: \url{https://github.com/isnowfy/snownlp}
\BIBentrySTDinterwordspacing

\bibitem{DBLP:journals/coling/TaboadaBTVS11}
M.~Taboada, J.~Brooke, M.~Tofiloski, K.~D. Voll, and M.~Stede, ``Lexicon-based
  methods for sentiment analysis,'' \emph{Comput. Linguistics}, vol.~37, no.~2,
  pp. 267--307, 2011.

\bibitem{DBLP:conf/icml/CollobertW08}
R.~Collobert and J.~Weston, ``A unified architecture for natural language
  processing: deep neural networks with multitask learning,'' in \emph{Machine
  Learning, Proceedings of the Twenty-Fifth International Conference {(ICML}
  2008), Helsinki, Finland, June 5-9, 2008}, ser. {ACM} International
  Conference Proceeding Series, vol. 307.\hskip 1em plus 0.5em minus
  0.4em\relax {ACM}, 2008, pp. 160--167.

\bibitem{article:wasserstein}
S.~S. Vallender, ``Calculation of the wasserstein distance between probability
  distributions on the line,'' \emph{Theory of Probability \& Its
  Applications}, vol.~18, no.~4, pp. 784--786, 1974.

\bibitem{article:kldivergence}
\BIBentryALTinterwordspacing
S.~Kullback and R.~A. Leibler, ``On information and sufficiency,'' \emph{The
  Annals of Mathematical Statistics}, vol.~22, no.~1, pp. 79--86, 1951.
  [Online]. Available: \url{http://www.jstor.org/stable/2236703}
\BIBentrySTDinterwordspacing

\bibitem{DBLP:journals/tit/Lin91}
J.~Lin, ``Divergence measures based on the shannon entropy,'' \emph{{IEEE}
  Trans. Inf. Theory}, vol.~37, no.~1, pp. 145--151, 1991.

\bibitem{DBLP:conf/nips/Cuturi13}
M.~Cuturi, ``Sinkhorn distances: Lightspeed computation of optimal transport,''
  in \emph{Advances in Neural Information Processing Systems 26: 27th Annual
  Conference on Neural Information Processing Systems 2013. Proceedings of a
  meeting held December 5-8, 2013, Lake Tahoe, Nevada, United States}, 2013,
  pp. 2292--2300.

\bibitem{DBLP:conf/msr/MartinHJSZ15}
W.~J. Martin, M.~Harman, Y.~Jia, F.~Sarro, and Y.~Zhang, ``The app sampling
  problem for app store mining,'' in \emph{12th {IEEE/ACM} Working Conference
  on Mining Software Repositories, {MSR} 2015, Florence, Italy, May 16-17,
  2015}.\hskip 1em plus 0.5em minus 0.4em\relax {IEEE} Computer Society, 2015,
  pp. 123--133.

\bibitem{DBLP:conf/chi/SrinivasanBLD18}
A.~Srinivasan, M.~Brehmer, B.~Lee, and S.~M. Drucker, ``What's the difference?:
  Evaluating variations of multi-series bar charts for visual comparison
  tasks,'' in \emph{Proceedings of the 2018 {CHI} Conference on Human Factors
  in Computing Systems, {CHI} 2018, Montreal, QC, Canada, April 21-26,
  2018}.\hskip 1em plus 0.5em minus 0.4em\relax {ACM}, 2018, p. 304.

\bibitem{DBLP:conf/chi/GrayKBHT18}
C.~M. Gray, Y.~Kou, B.~Battles, J.~Hoggatt, and A.~L. Toombs, ``The dark
  (patterns) side of {UX} design,'' in \emph{Proceedings of the 2018 {CHI}
  Conference on Human Factors in Computing Systems, {CHI} 2018, Montreal, QC,
  Canada, April 21-26, 2018}.\hskip 1em plus 0.5em minus 0.4em\relax {ACM},
  2018, p. 534.

\bibitem{DBLP:journals/corr/abs-2102-10772}
\BIBentryALTinterwordspacing
R.~Hu and A.~Singh, ``Transformer is all you need: Multimodal multitask
  learning with a unified transformer,'' \emph{CoRR}, vol. abs/2102.10772,
  2021. [Online]. Available: \url{https://arxiv.org/abs/2102.10772}
\BIBentrySTDinterwordspacing

\bibitem{DBLP:conf/ijcai/WangWZY17}
J.~Wang, Z.~Wang, D.~Zhang, and J.~Yan, ``Combining knowledge with deep
  convolutional neural networks for short text classification,'' in
  \emph{Proceedings of the Twenty-Sixth International Joint Conference on
  Artificial Intelligence, {IJCAI} 2017, Melbourne, Australia, August 19-25,
  2017}.\hskip 1em plus 0.5em minus 0.4em\relax ijcai.org, 2017, pp.
  2915--2921.

\bibitem{DBLP:journals/tpds/MiWZLC13}
H.~Mi, H.~Wang, Y.~Zhou, M.~R. Lyu, and H.~Cai, ``Toward fine-grained,
  unsupervised, scalable performance diagnosis for production cloud computing
  systems,'' \emph{{IEEE} Trans. Parallel Distributed Syst.}, vol.~24, no.~6,
  pp. 1245--1255, 2013.

\bibitem{DBLP:conf/sigsoft/GuWWZLKZYSXQLLZ20}
J.~Gu, J.~Wen, Z.~Wang, P.~Zhao, C.~Luo, Y.~Kang, Y.~Zhou, L.~Yang, J.~Sun,
  Z.~Xu, B.~Qiao, L.~Li, Q.~Lin, and D.~Zhang, ``Efficient customer incident
  triage via linking with system incidents,'' in \emph{{ESEC/FSE} '20: 28th
  {ACM} Joint European Software Engineering Conference and Symposium on the
  Foundations of Software Engineering, Virtual Event, USA, November 8-13,
  2020}.\hskip 1em plus 0.5em minus 0.4em\relax {ACM}, 2020, pp. 1296--1307.

\bibitem{DBLP:conf/icsm/PalombaVBOPPL15}
F.~Palomba, M.~L. V{\'{a}}squez, G.~Bavota, R.~Oliveto, M.~D. Penta,
  D.~Poshyvanyk, and A.~D. Lucia, ``User reviews matter! tracking crowdsourced
  reviews to support evolution of successful apps,'' in \emph{2015 {IEEE}
  International Conference on Software Maintenance and Evolution, {ICSME} 2015,
  Bremen, Germany, September 29 - October 1, 2015}.\hskip 1em plus 0.5em minus
  0.4em\relax {IEEE} Computer Society, 2015, pp. 291--300.

\bibitem{DBLP:conf/kdd/FuLLFHS13}
B.~Fu, J.~Lin, L.~Li, C.~Faloutsos, J.~I. Hong, and N.~M. Sadeh, ``Why people
  hate your app: making sense of user feedback in a mobile app store,'' in
  \emph{The 19th {ACM} {SIGKDD} International Conference on Knowledge Discovery
  and Data Mining, {KDD} 2013, Chicago, IL, USA, August 11-14, 2013}.\hskip 1em
  plus 0.5em minus 0.4em\relax {ACM}, 2013, pp. 1276--1284.

\bibitem{stradowski2023industrial}
S.~Stradowski and L.~Madeyski, ``Industrial applications of software defect
  prediction using machine learning: A business-driven systematic literature
  review,'' \emph{Information and Software Technology}, vol. 159, p. 107192,
  2023.

\bibitem{he2025practicability}
Z.~He, P.~Chen, and Z.~Zheng, ``On the practicability of deep learning based
  anomaly detection for modern online software systems: A pre-train-and-align
  framework,'' \emph{ACM Transactions on Software Engineering and Methodology},
  2025.

\bibitem{tan2023stre}
Y.~Tan, J.~Chen, W.~Shang, T.~Zhang, S.~Fang, X.~Luo, Z.~Chen, and S.~Qi,
  ``Stre: an automated approach to suggesting app developers when to stop
  reading reviews,'' \emph{IEEE Transactions on Software Engineering}, vol.~49,
  no.~8, pp. 4135--4151, 2023.

\bibitem{chen2024smart}
L.~Chen, Y.~Pei, M.~Wan, Z.~Fei, T.~Liang, and G.~Ma, ``Smart issue detection
  for large-scale online service systems using multi-channel data,'' in
  \emph{International Conference on Fundamental Approaches to Software
  Engineering}.\hskip 1em plus 0.5em minus 0.4em\relax Springer, 2024, pp.
  165--187.

\bibitem{DBLP:journals/infsof/FinkelsteinHJMS17}
A.~Finkelstein, M.~Harman, Y.~Jia, W.~J. Martin, F.~Sarro, and Y.~Zhang,
  ``Investigating the relationship between price, rating, and popularity in the
  blackberry world app store,'' \emph{Inf. Softw. Technol.}, vol.~87, pp.
  119--139, 2017.

\bibitem{DBLP:conf/bcshci/IacobVH13}
C.~Iacob, V.~Veerappa, and R.~Harrison, ``What are you complaining about?: a
  study of online reviews of mobile applications,'' in \emph{{BCS-HCI} '13
  Proceedings of the 27th International {BCS} Human Computer Interaction
  Conference, Brunel University, London, UK, 9-13 September 2013}.\hskip 1em
  plus 0.5em minus 0.4em\relax British Computer Society, 2013, p.~29.

\bibitem{DBLP:conf/kbse/VuNPN15}
P.~M. Vu, T.~T. Nguyen, H.~V. Pham, and T.~T. Nguyen, ``Mining user opinions in
  mobile app reviews: {A} keyword-based approach {(T)},'' in \emph{30th
  {IEEE/ACM} International Conference on Automated Software Engineering, {ASE}
  2015, Lincoln, NE, USA, November 9-13, 2015}.\hskip 1em plus 0.5em minus
  0.4em\relax {IEEE} Computer Society, 2015, pp. 749--759.

\bibitem{DBLP:conf/kbse/VuPNN16}
P.~M. Vu, H.~V. Pham, T.~T. Nguyen, and T.~T. Nguyen, ``Phrase-based extraction
  of user opinions in mobile app reviews,'' in \emph{Proceedings of the 31st
  {IEEE/ACM} International Conference on Automated Software Engineering, {ASE}
  2016, Singapore, September 3-7, 2016}.\hskip 1em plus 0.5em minus 0.4em\relax
  {ACM}, 2016, pp. 726--731.

\bibitem{DBLP:conf/issre/ManGLJ16}
Y.~Man, C.~Gao, M.~R. Lyu, and J.~Jiang, ``Experience report: Understanding
  cross-platform app issues from user reviews,'' in \emph{27th {IEEE}
  International Symposium on Software Reliability Engineering, {ISSRE} 2016,
  Ottawa, ON, Canada, October 23-27, 2016}.\hskip 1em plus 0.5em minus
  0.4em\relax {IEEE} Computer Society, 2016, pp. 138--149.

\bibitem{DBLP:conf/sigsoft/SorboPASVCG16}
A.~D. Sorbo, S.~Panichella, C.~V. Alexandru, J.~Shimagaki, C.~A. Visaggio,
  G.~Canfora, and H.~C. Gall, ``What would users change in my app? summarizing
  app reviews for recommending software changes,'' in \emph{Proceedings of the
  24th {ACM} {SIGSOFT} International Symposium on Foundations of Software
  Engineering, {FSE} 2016, Seattle, WA, USA, November 13-18, 2016}.\hskip 1em
  plus 0.5em minus 0.4em\relax {ACM}, 2016, pp. 499--510.

\bibitem{DBLP:conf/re/MaalejN15}
W.~Maalej and H.~Nabil, ``Bug report, feature request, or simply praise? on
  automatically classifying app reviews,'' in \emph{23rd {IEEE} International
  Requirements Engineering Conference, {RE} 2015, Ottawa, ON, Canada, August
  24-28, 2015}.\hskip 1em plus 0.5em minus 0.4em\relax {IEEE} Computer Society,
  2015, pp. 116--125.

\bibitem{DBLP:conf/icse/LuLLXMH0F16}
X.~Lu, X.~Liu, H.~Li, T.~Xie, Q.~Mei, D.~Hao, G.~Huang, and F.~Feng, ``{PRADA:}
  prioritizing android devices for apps by mining large-scale usage data,'' in
  \emph{Proceedings of the 38th International Conference on Software
  Engineering, {ICSE} 2016, Austin, TX, USA, May 14-22, 2016}.\hskip 1em plus
  0.5em minus 0.4em\relax {ACM}, 2016, pp. 3--13.

\bibitem{DBLP:conf/icdm/LimLZFTLDZ14}
M.~Lim, J.~Lou, H.~Zhang, Q.~Fu, A.~B.~J. Teoh, Q.~Lin, R.~Ding, and D.~Zhang,
  ``Identifying recurrent and unknown performance issues,'' in \emph{2014
  {IEEE} International Conference on Data Mining, {ICDM} 2014, Shenzhen, China,
  December 14-17, 2014}.\hskip 1em plus 0.5em minus 0.4em\relax {IEEE} Computer
  Society, 2014, pp. 320--329.

\bibitem{DBLP:conf/icse/LinLZZ16}
Q.~Lin, J.~Lou, H.~Zhang, and D.~Zhang, ``idice: Problem identification for
  emerging issues,'' in \emph{Proceedings of the 38th International Conference
  on Software Engineering, {ICSE} 2016, Austin, TX, USA, May 14-22,
  2016}.\hskip 1em plus 0.5em minus 0.4em\relax {ACM}, 2016, pp. 214--224.

\bibitem{DBLP:conf/icse/WuWCZ18}
R.~Wu, M.~Wen, S.~Cheung, and H.~Zhang, ``Changelocator: locate crash-inducing
  changes based on crash reports,'' in \emph{Proceedings of the 40th
  International Conference on Software Engineering, {ICSE} 2018, Gothenburg,
  Sweden, May 27 - June 03, 2018}.\hskip 1em plus 0.5em minus 0.4em\relax
  {ACM}, 2018, p. 536.

\bibitem{DBLP:conf/sigsoft/GuLQQL0LDCWZCZ20}
J.~Gu, C.~Luo, S.~Qin, B.~Qiao, Q.~Lin, H.~Zhang, Z.~Li, Y.~Dang, S.~Cai,
  W.~Wu, Y.~Zhou, M.~Chintalapati, and D.~Zhang, ``Efficient incident
  identification from multi-dimensional issue reports via meta-heuristic
  search,'' in \emph{{ESEC/FSE} '20: 28th {ACM} Joint European Software
  Engineering Conference and Symposium on the Foundations of Software
  Engineering, Virtual Event, USA, November 8-13, 2020}.\hskip 1em plus 0.5em
  minus 0.4em\relax {ACM}, 2020, pp. 292--303.

\bibitem{DBLP:conf/icse/LinZLZC16}
Q.~Lin, H.~Zhang, J.~Lou, Y.~Zhang, and X.~Chen, ``Log clustering based problem
  identification for online service systems,'' in \emph{Proceedings of the 38th
  International Conference on Software Engineering, {ICSE} 2016, Austin, TX,
  USA, May 14-22, 2016 - Companion Volume}.\hskip 1em plus 0.5em minus
  0.4em\relax {ACM}, 2016, pp. 102--111.

\bibitem{DBLP:conf/sigsoft/HeLLZLZ18}
S.~He, Q.~Lin, J.~Lou, H.~Zhang, M.~R. Lyu, and D.~Zhang, ``Identifying
  impactful service system problems via log analysis,'' in \emph{Proceedings of
  the 2018 {ACM} Joint Meeting on European Software Engineering Conference and
  Symposium on the Foundations of Software Engineering, {ESEC/SIGSOFT} {FSE}
  2018, Lake Buena Vista, FL, USA, November 04-09, 2018}.\hskip 1em plus 0.5em
  minus 0.4em\relax {ACM}, 2018, pp. 60--70.

\end{thebibliography}
\balance


\end{document}